\documentclass[a4paper,11pt]{article}
\pdfoutput=1
\usepackage{jinstpub}
\usepackage{hyperref}

\title{Realistic Monte Carlo simulations of silicon 4D-trackers}

\author{M. Mandurrino}
\affiliation{INFN -- Sezione di Torino,\\Via P. Giuria, 1 -- 10125 Torino, Italy}
\emailAdd{marco.mandurrino@to.infn.it}

\abstract{Simulation-guided design represents a fundamental contribution towards the development of modern semiconductor devices aiming to reach high-performance particle detection, identification and tracking, and constitutes a strategic element of the new detector R\&D roadmap.
At the same time, the complexity of microelectronic structures and the related detection systems is drastically increasing, also thanks to the progressive scaling down of the design rules with the process technology.
Owing to the capability to embed a detailed description of the ionization mechanism into a device-level framework, as well as capture the stochastic nature of signal formation, the Monte Carlo (MC) approach has become the most recommended strategy to achieve reliable predictions of the dynamic properties of particle detectors in realistic settings such as in-beam experiments.
This work gives an overview of the key aspects characterizing MC tools, with particular emphasis on the Garfield++ simulation toolkit.
To this end, the analysis of some specific case studies related to the design of silicon particle detectors for timing and 4D-tracking in both current and future high-energy physics experiments will be presented, showing the comparison of measured and simulated figures-of-merit and highlighting strengths and open challenges of this approach.
The examples are intentionally chosen from the family of Monolithic Active Pixel Sensors, as they represent some of the most promising and relevant advancements in particle detection, and because the CMOS monolithic integration offers the most versatile platform for testing the robustness of numerical designs.}


\proceeding{11$^{\text{th}}$ International Workshop on Semiconductor Pixel Detectors for Particles and Imaging (PIXEL2024)\\
  18-22 November 2024\\
  Strasbourg}

\begin{document}
\maketitle
\flushbottom

\graphicspath{{f./}}

\section{Introduction}
\label{sec:intro}
As the technological requirements of high-energy physics experiments become increasingly stringent and the complexity of solid-state particle detectors grows, it is necessary that numerical simulation, a crucial element in sensor development, becomes a robust and reliable design tool.
If until about a decade ago deterministic techniques, such as the Finite Element Method (FEM), could be sufficient, the advent of sensors integrated monolithically with the readout electronics, and the need for elevated performance in terms of timing and tracking precision, has made the use of stochastic methods for modelling charge transport in semiconductors a necessary step.
To this end, Monte Carlo (MC) algorithms began to be employed in the simulation-driven design of particle detectors.
The first examples, dating back to the early 1980s, were conceived to model the signal formation in gaseous detectors -- such as drift chambers -- starting from electric field distribution maps~\cite{garfield}.
Then, nearly three decades later, the same approach was extended to the simulation of semiconductor devices~\cite{garfieldpp,allpix2}, thanks to the development of a set of software frameworks customized for silicon-based particle detectors.
Nevertheless, since MC simulations require the field maps generated by FEM tools, the Technology Computer-Aided Design (TCAD) approach remains an essential part of the design process, as it offers the possibility to optimize numerical models and calibrate simulation parameters against technology parameters.

In this work, the main features of one of the most widely used MC tools for simulating particle detectors, \emph{Garfield}++, are presented, with a specific focus on a real application case related to the design of monolithic detectors for high-precision particle track reconstruction in four dimensions.
After introducing the software, the development of high-performance 4D trackers relying on realistic numerical simulations will be presented.
The studies that led to the development of the first CMOS-LGAD detectors for timing, fabricated in the 110 nm technological node~\cite{2023Follo_NIMA}, will be initially discussed.
Then, the numerical proof-of-concept for the monolithic integration of the RSD paradigm~\cite{2020Mandurrino_NIMA} into a standard CMOS process will be finally detailed.

\section{Monte Carlo simulation of solid-state particle detectors}
\label{sec:montecarlo}
As stated on the reference webpage, \emph{Garfield}++ (or simply G++ hereafter) is an object-oriented toolkit for the detailed simulation of particle detectors based on ionization measurement in gases or semiconductors.
It is a direct evolution of  \emph{Garfield}, the predecessor tool developed by Rob Veenhof in 1984~\cite{garfield}, retaining much of its functionality but introducing several improvements, such as the user interface, that now is ROOT-based, and a new treatment for charge carrier transport.
The most important features of G++ include (\emph{i}) the ionization pattern simulation, suitable for relativistic particles, X-ray absorption, low-energy ions and other ionizing particles, (\emph{ii}) the transport mechanisms, available for both gas mixtures and semiconductors, and (\emph{iii}) advanced electric field calculation methods.
Relativistic particles and X-ray photoabsorption simulations share an interface with the HEED (High-Energy ElectroDynamics) program~\cite{2005Smirnov_NIMA}, which is based on the PAI (\mbox{Photo-Absorption} Ionization) method~\cite{1980Allison_ARNPS}.
Ion tracks are calculated using external tools such as SRIM (Stopping Range of Ions in Matter) or TRIM (TRansport of Ions in Matter).
For other types of simulations, interfaces with Magboltz~\cite{magboltz} and GEANT4 are available to calculate ionization/avalanche in gases and particle tracks in generic media, respectively.

As for the charge transport, the second innovative aspect characterizing G++, there are at least two different strategies to solve the first-order equation of motion and compute the drift lines of each individual electron and hole: the Runge-Kutta-Fehlberg (RKF) integration, which iteratively estimates charge position, drift velocity and time step, and the Monte Carlo (MC) integration -- which is what we are interested in -- based on macroscopic transport parameters.
Following the MC method, a step of length $\Delta s = v_\textrm{d} \Delta t$ along the drift velocity $v_\textrm{d}$ direction is calculated after the user specifies the time step $\Delta t$ (the same calculation can be done for $\Delta t$, initializing $\Delta s$).
Then, a random diffusion step is generated, and the new charge carrier location is updated by vectorially adding the two steps.
The random diffusion step is sampled from three uncorrelated Gaussian distributions with standard deviations $\sigma = D \sqrt{\Delta s}$, where the coefficient $D$ consists of one component parallel to the drift velocity and two transverse components.

The third key element distinguishing G++ is the electric field calculation.
To simplify, there are two main procedures: importing the field as an external input file or calculating it directly.
In the first case, the input can consist of simple text files listing both the field intensity and the corresponding coordinates.
Alternatively, complex field (and potential) maps can be generated using FEM tools, such as Ansys, TCAD, COMSOL, and other software programs.
Notably, G++ does not require the generation of a new discretization of the geometry, as it can adapt to the one created by the external tools, when generating the maps.
When directly computing the electric field -- this approach is only valid when the device geometry is simple enough for the field shape to be predicted analytically -- several interfaces are available, including neBEM (nearly-exact Boundary Element Method) or the thin-wire limit solution.

Once the input field is produced or imported, the user must perform the following steps to obtain the current signals generated by the passage of an ionizing particle through the detector: (\emph{a}) calculate the weighting field and weighting potential maps, (\emph{b}) activate the desired physical models, (\emph{c}) choose the particle type, its momentum (or energy) and trajectory and, finally, (\emph{d}) simulate the signal formation.
Regarding step (\emph{a}), to convert the static potential $\Psi$ (field $\mathbf{E}=-\nabla \Psi$) into the static weighting potential $\Psi_\textrm{w}$ (field $\mathbf{E}_\textrm{w}=-\nabla \Psi_\textrm{w}$), it is necessary to compute or import the electrostatic maps in two conditions: before and after a potential step $\delta V$ is applied to the readout electrode, then:
\begin{equation}
\Psi_\textrm{w} = \frac{\Psi (V+\delta V) - \Psi (V)}{\delta V} \, .
\end{equation}
Typically, a $\delta V$ equal to 1\% of the bias $V$ is sufficient.
Furthermore, relevant for the implementation of point (\emph{b}) is the activation of proper impact ionization models when the detector is based on the internal multiplication mechanism.
In this regard, G++ offers four different options to model the ionization coefficients, all based on the Chynoweth functional form~\cite{1958Chynoweth_PR}: the Grant~\cite{1979Grant_SSE}, Massey~\cite{2006Massey_TED}, van Overstraeten-de Man~\cite{1970Overstraeten_SSE} and the Okuto-Crowell~\cite{1975Okuto_SSE} model.
Finally, step (\emph{c}) is rather straightforward, while the last point -- signal formation -- requires one final set of input information about the transient granularity that the user must initialize.
This includes the initial time, final time, and the number of bins in which the signal is acquired by the readout electrode.

When the device is defined, all the preliminary steps are completed and the particle has crossed the detector, the transport of charge carriers is activated as previously described and the induced current $i(t)$ is calculated through the Shockley-Ramo theorem~\cite{1938Shockley_JAP,1939Ramo_PIRE}:
\begin{equation}
i(t) = -q \, \dot{\mathbf{x}}(t) \cdot \mathbf{E}_\textrm{w} \left(\mathbf{x}(t)\right)  \, ,
\end{equation}
where $q$ is the charge of a carrier moving at position $\mathbf{x}$ and velocity $\dot{\mathbf{x}}$.

Before analysing the simulation of a real particle detector, it is worth mentioning that G++ allows the inclusion of the contribution from the readout electronics to the final shape of the signal. This can be achieved by convoluting the induced current with a transfer function $f(t)$, which can be defined through a user-specific function, a table of values or one of the pre-implemented analytic models.
One of the most commonly used models is the \mbox{\textit{n}-stage} shaper, where
\begin{equation}
f(t) = g \, \mathrm{exp}(n) \left(t / t_p\right)^n \mathrm{exp}(-t/\tau) \, ,
\end{equation}
with $t_p = n \tau$.
In addition, also a source of noise can be added, reproducing a given equivalent-noise charge at the amplifier output.

\section{Simulating monolithic LGADs and AC-LGADs}
\label{sec:simulations}
\begin{figure}[htbp]
\centering
\includegraphics[width=.55\columnwidth]{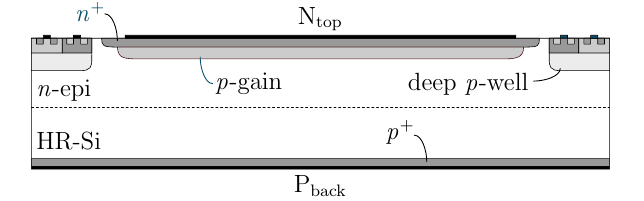}
\caption{\label{fig:cmoslgad} Simplified cross-section of a CMOS-LGAD detector.}
\end{figure}

The goal of this section is to show how MC simulations can be employed to design and optimize a particle detector intended for a specific experiment or application.
The use case here reported is the monolithic LGAD detector in 110 nm CMOS technology, under development at INFN Torino within the ARCADIA project, for timing measurements in high-energy physics experiments.
The cross-section of the device, which is based on a fully-depleted Monolithic Active Pixel Sensor (FD-MAPS) structure, is reported in figure~\ref{fig:cmoslgad}.
Starting from the bottom, there is a \textit{p}$^+$-type region, working as a back-side contact.
Grown on a \textit{n}-type high-resistivity (HR) substrate, we find an epitaxial layer, with the same polarity of the substrate but higher doping, to delay the onset of punch-through current.
This current consists of holes flowing between the back-side \textit{p}$^+$-contact (P$_\textrm{back}$) and the deep \textit{p}-well on the front-side.
The substrate and the epitaxial layer constitute the detector active volume.
Finally, the top of the device hosts the sensing region -- consisting of a \textit{n}$^+$ collection electrode and the underlying \textit{p}-type gain layer -- as well as the readout electronics -- represented here by NMOS and PMOS transistors within the lateral deep \textit{p}-wells.
While the bias on N$_\textrm{top}$ controls the intensity of the electric field in the sensing region and, thus, the multiplication factor (40-60~V may be required to generate a gain in the range of 15 to 30), the reverse voltage applied to P$_\textrm{back}$ mainly determines the depletion of the active volume.
Typically, 20-30~V are sufficient to achieve full-depletion (FD) in a 50-$\mu$m-thick detector.
If this reverse polarization goes beyond a certain value, a conductive channel creates between the \textit{p}-type implants on the top- and back-side, and a punch-through (PT) current starts to flow.
Consequently, the CMOS-LGAD operates in the window $V_\textrm{FD}<V_\textrm{back}<V_\textrm{PT}$, where $V_\textrm{PT}$ depends on the \textit{n}-epi thickness and doping, as well as on the top-side bias.

After producing two lots of standard MAPS detectors, where the principal sensor figures-of-merit have been extensively studied, structures with internal multiplication have been included in the ARCADIA third engineering run.
To design and optimize CMOS-LGADs, a numerical MC framework has been developed.
Of course, such a tool needs to be optimized and calibrated with respect to the new manufacturing parameters, especially those related to the implementation of a new implant: the gain layer.
Additionally, to study the impact of the periphery on sensor operation, two different layouts have been implemented: A1, where the lateral deep \textit{p}-wells extend until they contact the gain layer, and A2, in which there is a gap between them (as shown in the cross-section of figure~\ref{fig:cmoslgad}).
In the first case, all the drift lines, even those coming from the sensor periphery, must cross the gain layer before reaching the collection electrode.
This means that all the charges generated in the active volume are multiplied.
However, signals from particles traversing the sensor under the lateral \textit{p}-wells are delayed relative to those produced at the center, due to the longer field lines.
As a result, multiplication is uniform, but the overall timing resolution is slightly degraded.
In case A2, on the other hand, some drift lines pass through the gap and reach the \textit{n}$^+$ cathode without undergoing charge multiplication.
Consequently, within the active area, the timing response is more uniform than in A1.

Before presenting some examples of numerical characterizations obtained by simulating with G++ the device shown in figure~\ref{fig:cmoslgad}, a preliminary calibration is provided in the next section.
We begin with the optimizaton of the numerical parameters, which is necessary to achieve a self-consistent description, followed by the extraction of key aspects from the matching between measured and simulated electrical characteristics, essential for establishing a realistic simulation framework.
Since the potential (field) maps generated with TCAD serve as the input for the MC simulation, it is important to ensure that they reflect -- as realistically as possible -- the material properties and the physical operation of the device under investigation.
To this end, preliminary tests have been conducted to study the extent to which TCAD and MC signals compare in simplified structures.

\begin{figure}[htbp]
\centering
\includegraphics[width=.75\columnwidth]{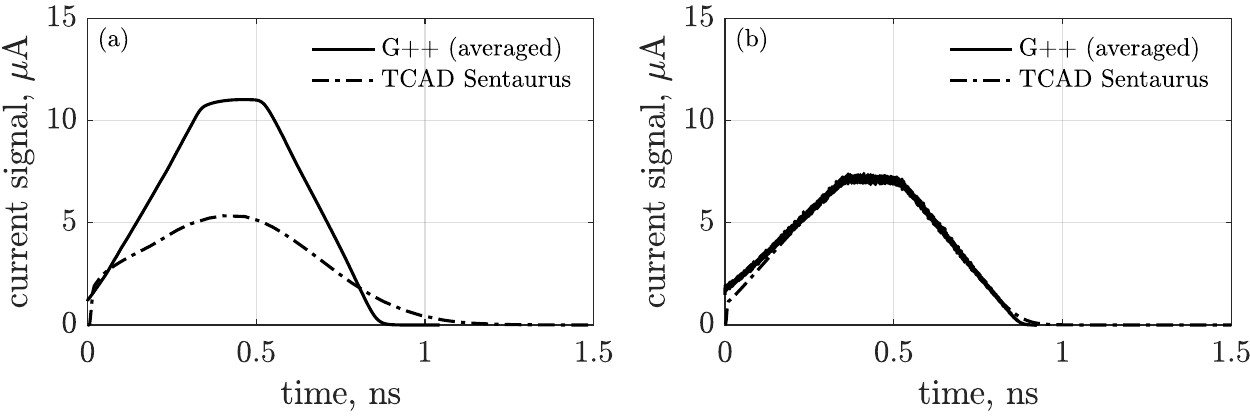}
\caption{\label{fig:signals} Comparison of MIP signals simulated with G++ (solid) and TCAD (dash-dot) in a 50-$\mu$m-thick  LGAD diode: (a) before and (b) after parameter calibration.}
\end{figure}

In figure~\ref{fig:signals} the effects of tuning some important physics models and parameters are shown by comparing the MIP signals produced by a 50-$\mu$m-thick  LGAD diode, as calculated with G++ and TCAD Sentaurus in a 2D framework.
Panel (a) shows the signals before optimization, and panel (b) after.
In the left plot, the default settings are used for both simulation tools, which also include the van Overstraten-de Man model for the impact ionization process.
The signals in G++ are obtained by injecting 180 GeV pions, with all the waveforms (several thousands) being averaged.
In TCAD Sentaurus, on the other hand, a single event is simulated using the heavy-ion model~\cite{Sentaurus}.

By adjusting in G++ the time step to $0.1$~ps, setting the dimension along the third axis to $10~\mu$m, and selecting, in TCAD Sentaurus, the extended Canali model for mobility, along with setting an ion track width of $3~\mu$m, a transient maximum step of $0.5$~ps, and reducing the number of electron/hole pairs produced per micron, it is possible to recover the mismatch shown in panel (a) and achieve the excellent agreement reported in panel (b).

In order to enable a consistent comparison between the TCAD and Monte Carlo simulations, the screening effects, naturally included in the TCAD approach but absent in the MC formalism, were deliberately mitigated in the TCAD simulation.
This was done not because such effects are negligible, but to align the physical assumptions of both methods.
Specifically, the mitigation was applied only to the primary charge carriers in the TCAD simulation by reducing the number of generated electron-hole pairs per micron.
To preserve the total induced signal and ensure \mbox{self-consistency} among the tools, the TCAD current was subsequently scaled by a constant factor.
No modification was introduced in the MC simulation.

Apart from selecting a specific value for numerical variables or choosing a particular physics model, what deserves close attention is the parameter that controls energy transfer in the ionization process.
Usually, TCAD tools take into account screening effects, and their time evolution, that arise when a large number of charge carriers are close to each other.
In contrast, MC simulations, which rely solely on the static field/potential map as input, do not consider these effects.
This appears to be the most reasonable explanation for why the number of pairs must be decreased (and the current consistently multiplied) in order to mitigate space charge effects in TCAD and compensate for the mismatch with respect to the MC signal.
\begin{figure}[htbp]
\centering
\includegraphics[width=.59\columnwidth]{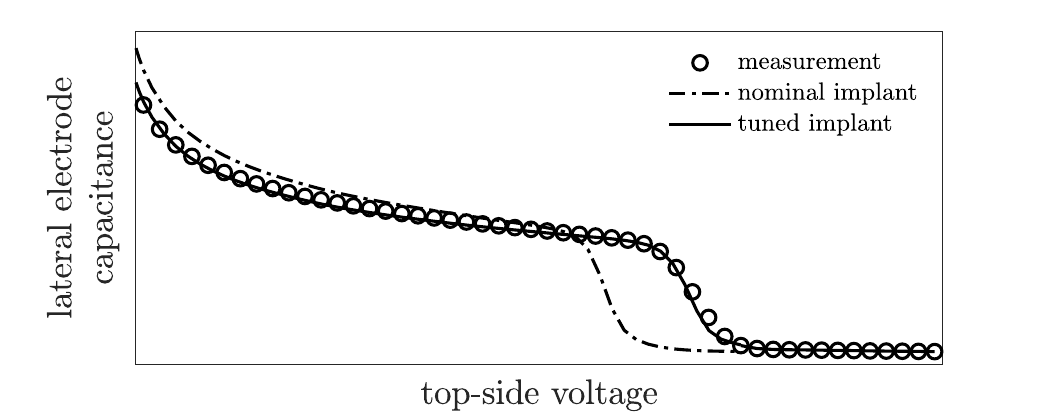}
\caption{\label{fig:cv} Lateral $C(V)$ characteristics of the collection electrode, measured (circles) and simulated with TCAD Sentaurus using the nominal (dash-dot) and tuned (solid) \emph{p}-gain implant profiles.}
\end{figure}
Besides the self-consistency among tools, technological aspects also play a central role in conducting realistic simulations.
To extract as much information as possible about the newly added \textit{p}-gain implant, the TCAD simulations were calibrated using the electrical characteristics measured in sensors from the third ARCADIA batch.
Specifically, as shown in figure~\ref{fig:cv}, the lateral capacitance of the collection electrode was compared with the corresponding simulation performed using the nominal implantation profile provided by the foundry.
As can be seen, the nominal profile does not correctly reproduce the experimental $C(V)$ unless fluctuations in the \textit{p}-gain implantation energy are taken into account, as is the case for the curve labeled `tuned implant'.
All the following simulations are optimized based on the previous considerations regarding self-consistency between tools and the calibration of technological parameters with measurements.

\begin{figure}[htbp]
\centering
\includegraphics[width=.75\columnwidth]{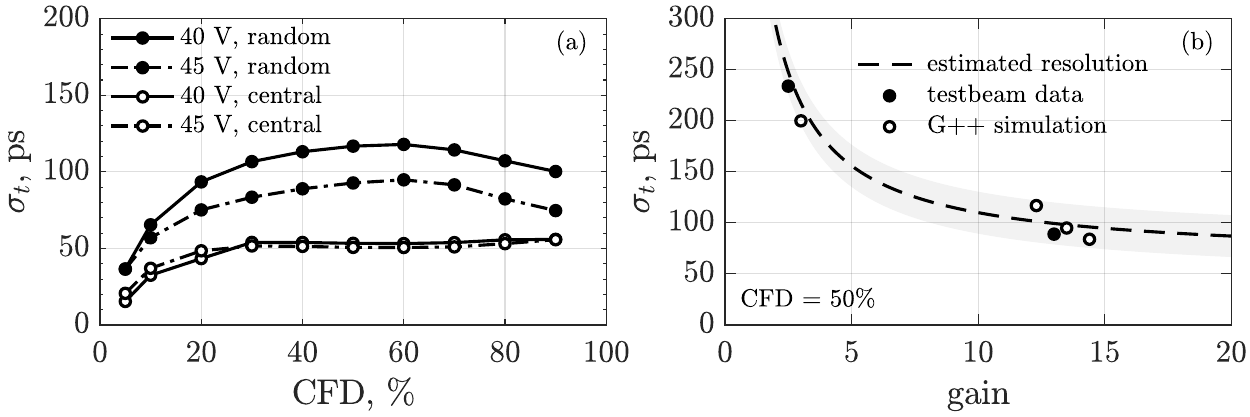}
\caption{\label{fig:sigma} CMOS-LGAD simulated time resolution as a function of (a) constant fraction discriminator and (b) sensor internal gain.}
\end{figure}
As CMOS-LGADs are being developed for timing purposes, the next result focuses on time resolution.
Using G++ to simulate 180 GeV/c pions impinging on the detector shown in figure~\ref{fig:cmoslgad}, the time resolution as a function of signal amplitude fraction (constant fraction discriminator, CFD) is calculated and shown in panel (a) of figure~\ref{fig:sigma} for two different biases of N$_\textrm{top}$ ($40$ and $45$~V).
The relativistic particles have been injected according to two configurations: point-like injection at center and random irradiation across a window covering the whole device.
Although the layout is A2, which is expected to be the most suitable for timing, the resolution is acceptably low only with particles at the center ($\sim$$50$~ps), while it increases remarkably in the random configuration (almost doubling for $V_\textrm{top}=40$~V).
The poor resolution and different behavior between the two types of simulation can be attributed to two main factors.
First, the periphery plays a crucial role, since non-multiplied signals degrade the overall resolution.
Second, there is an electrostatic effect observed in the TCAD input maps.
Specifically, a field sag beneath the gain layer slows down the charges, resulting in poorer time resolution.
This effect arises from a non-optimized substrate and low top-side bias.
Increasing $V_\textrm{top}$ from $40$ to $45$~V, indeed, significantly improves the time resolution, particularly in the random configuration.

Being focused on providing realistic numerical predictions of detector performance, in panel (b) of figure~\ref{fig:sigma} we compared the simulated (in the random configuration) and measured time resolutions versus gain of a \mbox{CMOS-LGAD} from the 3rd ARCADIA engineering run, calculated at $\textrm{CFD}=50\%$.
The dashed line represents the fit curve obtained by interpolating the simulations (open circles), with the grey area indicating the confidence interval.
Black dots correspond to the test beam data obtained in October 2023 ($\textrm{gain}=2.5$) and 2024 ($\sim$$13$).
The agreement is quite satisfactory, suggesting that both the calibration procedure and the G++ simulations are robust and reliable.

\begin{figure}[htbp]
\centering
\includegraphics[width=.99\columnwidth]{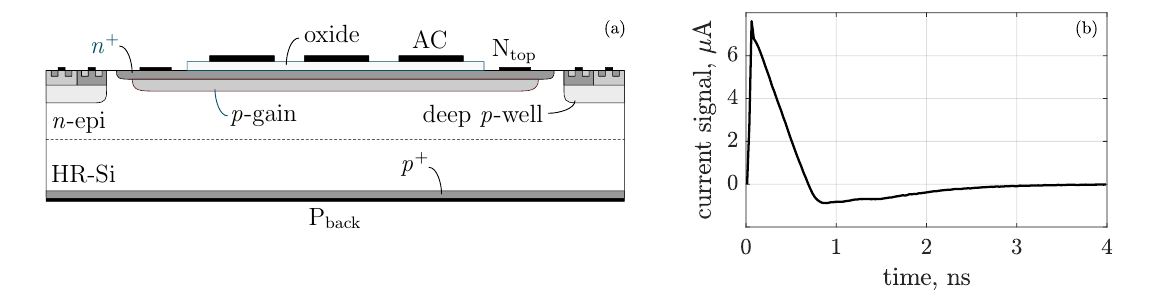}
\caption{\label{fig:cmosaclgad} Simplified cross-section of a CMOS-AC-LGAD detector (a) and waveform (average) simulated with G++ by injecting a 180 GeV/c pion in the middle of a metal pad.}
\end{figure}

In the last part of this work we present the numerical characterization of an innovative \mbox{LGAD-based} monolithic detector designed for 4D particle tracking.
This device (see figure~\ref{fig:cmosaclgad}) derives from the combination of two paradigms, the well-known RSD technology~\cite{2020Mandurrino_NIMA} and the CMOS process for circuit integration.
To account for the transport properties of the resistive electrode on which RSDs are based, the extended Shockley-Ramo theorem~\cite{2004Riegler_NIMA} implemented in G++ has been used.
This formalisms relies on a modified definition of the weighting potential that includes a static (prompt) and a time-dependent (delayed) component.
The new dynamic weighting potential, thus, becomes $\Psi_\textrm{w}(\mathbf{x},t) = \Psi_\textrm{w}^\textrm{p}(\mathbf{x}) + \Psi_\textrm{w}^\textrm{d}(\mathbf{x},t)$.
The first component, $\Psi_\textrm{w}^\textrm{p}(\mathbf{x})$, coincides with the static potential $\Psi_\textrm{w}$, while calculating the delayed term $\Psi_\textrm{w}^\textrm{d}(\mathbf{x},t)$ requires sampling the transient that follows the application of the potential step to the readout electrode.
According to the extended Shockley-Ramo theorem, the expression for the induced current is given by
\begin{equation}
i(t) = -q \, \dot{\mathbf{x}}(t) \cdot \mathbf{E}_\textrm{w} \left(\mathbf{x}(t)\right) -q \int_0^t \mathbf{H}_\textrm{w} \left(\mathbf{x}(t^\prime),t-t^\prime\right)
\cdot \dot{\mathbf{x}}(t^\prime) \, \mathrm{d}t^\prime \, ,
\end{equation}
where the first term represents the standard direct induction, the second one describes the reaction from resistive medium, and $\mathbf{H}_\textrm{w}=-\nabla (\partial \Psi_\textrm{w}(\mathbf{x},t) / \partial t)$ is the dynamic weighting field.
Using this formalism, G++ was employed to generate signal pulses, like the one reported in panel (b) of figure~\ref{fig:cmosaclgad}, to test the functionality of the innovative CMOS-AC-LGAD.
The waveforms obtained show the same figures-of-merit as those produced by standard RSDs, namely: (\emph{i}) bipolar behavior, due to the capacitive coupling of the multiplied charges with the AC metal pads; (\emph{ii}) a prominent main lobe (large and fast induction), which is compatible with high performance in terms of timing and proportional to the distance between the hit point and the pad; and (\emph{iii}) a long, less pronounced, second lobe (slow discharge).
These characteristics suggest strong alignment with high spatial resolution and a 100\% fill factor, typical of the RSD paradigm.

\section{Conclusions}
\label{sec:conclusions}
This work discusses the role of Monte Carlo (MC) simulations in designing high-performance particle detectors, particularly in high-energy physics experiments.
As detector complexity increases, stochastic methods, such as MC algorithms, have become essential for modeling charge transport in semiconductors.
After focusing on the MC toolkit \emph{Garfield}++, simulations of CMOS-LGADs and AC-LGADs have been discussed, illustrating the importance of calibrating numerical parameters and technological aspects for accurate predictions of detector performance.

\acknowledgments
This project has received funding from the European Union’s Horizon 2020 Research and Innovation programme, under AIDAInnova G.A. 101004761, and from the Italian National Institute for Nuclear Physics (INFN) within the CSN5 call ARCADIA.

\end{document}